# Analysis of Applicability of ISO 9564 PIN based Authentication to Closed-Loop Mobile Payment Systems

**Amal Saha**[*]

Tata Institute of Fundamental Research (TIFR), Mumbai, INDIA,
Email: amal.k.saha@gmail.com

**Sugata Sanyal**

Tata Consultancy Services (TCS), Mumbai, INDIA
Email: sugata.sanyal@tcs.com

*Corresponding Author

—————————————————Abstract———————————————————

Payment transactions initiated through a mobile device are growing and security concerns must be addressed. People coming from payment card industry often talk passionately about porting ISO 9564 PIN standard based authentication in open-loop card payment to closed-loop mobile financial transactions and certification of closed-loop payment product or solution against this standard. In reality, so far this standard has not been adopted in closed-loop mobile payment authentication and applicability of this ISO standard must be studied carefully before adoption. The authors do a critical analysis of the applicability of this ISO specification and makes categorical statement about relevance of compliance to closed-loop mobile payment. Security requirements for authentication in closed-loop mobile payment systems are not standardised through ISO 9564 standard, Common Criteria [3], etc. Since closed-loop mobile payment is a relatively new field, the authors make a case for Common Criteria Recognition Agreement (CCRA) or other standards organization to push for publication of a mobile device-agnostic Protection Profile or standard for it, incorporating the suggested authentication approaches.

## I. INTRODUCTION

Financial inclusion in developing economies has been possible with emergence of mobile closed-loop stored-value-account based payment system that provides an alternative to standard banking system. In African and Asian countries and also in other emerging economies, closed-loop payment systems have established significant footprint. M-PESA [23] in Africa has been a huge success.

Open-loop payment networks such as Visa, MasterCard connect multiple parties —cardholder, financial institution that issues payment cards, a.k.a. issuer, the merchant and the one that has banking relationship with the merchants, a.k.a. acquirer— and manages flow of financial data and processing among them. Closed-loop payment network operator provides payment services directly to merchants and cardholders, without involving third-party financial institution as intermediary.

Mobile closed-loop payment applications enable consumers to manage and usestored-value account (SVA) and digital wallet, through their mobile device. This is used, for making payment, using backend services and also using POS terminals in some cases. Various communication channels like USSD, SMS, HTTP(S), etc may be used between application or component on mobile device and the backend services. A basic closed-loop mobile payment system may be functionally described as a hierarchical money distribution and payment system involving agents and whole-sale distributor and also mobile subscribers, with an accounting system that supports stored-value (SVA), i.e., prepaid system. USSD and SMS are often used as channels and a basic or feature phone is sufficient. As a result, authentication needed for initiation of closed-loop transactions by subscribers is often based on mobile PIN (m-PIN).

For online purchase using laptops, desktops and smartphones, device identification [6, 13] techniques have been used as additional authentication mechanism. Hristo Bojinov et al[16] demonstrated how the multitude of sensors on a smartphone can be used to construct a reliable hardware fingerprint of the phone and how such a fingerprint can be used to de-anonymise mobile devices as they connect to web sites, and as a second factor in identifying legitimate users to a remote server. Michael Rausch et al [18] observed cookies are the dominant technology for tracking user on the internet, some companies have developed an alternative form of tracking known as device fingerprinting that does not rely on cookies to identify a visitor. In this method, a profile of the user may be created by querying the browser through JavaScript about characteristics such as browser version, screen size, fonts installed and more [17, 18, 19].

It is assumed that we do not want to couple the solution of hardening the mobile PIN (m-PIN) with the SIM (UICC) or other forms of Secure Element [24] provided by the telecom operator and m-PIN may be used for various access bearers like USSD, SMS, NFC (near field communication) and mobile handset apps. Mobile devices have been used for authentication in web payment since many years [14, 15] and general concept of authentication in the context of mobile is well established.

In this paper, the focus is on applicability of ISO 9564 PIN based authentication to specific type of closed-loop payment system where payment is initiated from an application running on device operating system and where the owner of the closed-loop payment system is typically a mobile network operator or a bank.

## II. LITERATURE SURVEY

### A. Review of ISO 9564 [1,2] Card PIN Management in Open-Loop Card Payment Authentication

I. In ISO 9564, card PINs are protected by a secure PIN block, to prevent dictionary attacks.

II. To protect the PIN during transmission from the secure PIN entry device (e.g., a component in the POS terminal) to the verifier, the standard requires that the PIN be encrypted, and specifies several formats that may be used. In each case, the PIN is encoded into a 64-bit PIN block, which is then encrypted by an algorithm approved by the specification.

III. For online interchange transactions, PINs must be protected by encryption using ISO 9564–1 PIN-block formats 0, 1, or 3 and also using physical security features of secure PIN entry devices. Format 3 is recommended.

a. For ISO format 0 and 3, the clear-text PIN block and the Primary Account Number (PAN) block must be XOR'ed together and then Triple-DES encrypted in electronic code book (ECB) mode to form the 64-bit output cipher block (the reversible encrypted PIN block).

b. ISO format 1 and format 2 are formed by concatenation of two fields: the plain-text PIN field and the filler field.

c. The PIN entry device shall include tamper-detection and response mechanisms which, if attacked, cause the PIN Entry Device (PED) [10] to become immediately inoperable and result in the automatic and immediate erasure of any secret information that might have been stored in the PED, such that it becomes infeasible to recover the secret information. The PIN entry device should be able to authenticate itself to the acquirer such that, once compromised, it is no longer able to authenticate itself to the acquirer.

IV. ISO 9564 PIN block format 2 must be used for encrypting card PINs that are submitted from the IC card reader to the IC card (a fact known as offline authentication).

V. All cardholder PINs processed offline using Integrated Circuit (IC) card technology must be protected in accordance with the requirements in Book 2 of the EMV IC Card Specifications for Payment Systems and ISO 9564.

VI. The provisions of ISO 9564-1:2011 are not intended to cover PIN management and security in environments where no persistent cryptographic relationship exists between the transaction-origination device and the acquirer, e.g., use of a browser for payment in online shopping.

VII. Persistent cryptographic relationship of the type mentioned above may be established by sharing a secret, e.g., symmetric encryption key between transaction originating secure cryptographic device (component in ATM, POS terminals, etc) and the acquirer processor.

VIII. Secure cryptographic devices used in ISO 9564 compliant systems are typically FIPS 140-2 Level 3 and EMVCo Level1 and Level2 [10, 11, 12] certified. Most mobile devices in the market are not FIPS 140-2 certified and high-end mobile devices in the market which are a very small fraction of overall mobile devices in circulation are certified at FIPS 140-2 Level 1 only.

## B. QR Code for Closed-Loop Payment

Camera of mobile device of the customer, along with software, may scan payment information provided by merchant in the form of QR Code and make payment. In case of closed-loop payment, an application on mobile device would authenticate the mobile device and/or the customer with the backend service. Authentication may be based on m-PIN and optionally some form of device identity or device fingerprinting [6, 16]. Advantage is that most phones can scan QR code, but number of phone with NFC capability is still a small fraction.

## C. Closed-Loop NFC Payment, Android HCE [9] and EMVCo Specification [4]

NFC channel has also been used in closed-loop payment where an application on mobile device operating system or secure element (SE) (e.g., SIM card, embedded SE, etc) on an NFC-enabled device used by the customer, creates cryptogram to interact with NFC-enabled closed-loop POS-terminal. This process directly debits fund from the stored value account of the account holder. Here too, authentication is needed and often it is in the form of an m-PIN that is typically entered on the NFC-enabled mobile device of the customer and is sent to the backend service of the closed-loop payment system through the application on the customer's device. This is one variant of closed-loop NFC payment. Some implementation used SIM card of the mobile device as Secure Element and the m-PIN entered by the customer may be validated offline in Secure Element and the PIN is often used for authorising cryptographic operation inside the Secure Element.

In NFC card emulation mode [7], a mobile device can emulate contactless smart card(such as those used for contactless payments, transit fare payment and building, etc) when tapped on a contactless reader or point-of-sale (POS) terminal. Until recently, the emulated contactless card application has been stored in a secure element (SE) [7, 8], defined by Global Platform as a tamper-resistant smart card capable of securely hosting applications and their confidential and cryptographic data. With introduction of host card emulation (HCE) [9],it is now possible to store the emulated contactless card application as a service running locally on the mobile device operating system (Service is a type of Application Component as per Android application framework [20] ). Google's Android OS (v4.4 "KitKat"onwards) and the Blackberry OS support this.NFC controller in mobile device can route communication from the contactless reader or POS terminal to an HCE service on the mobile device.

With HCE, the service running on the mobile host operating system can interface with a contactless reader or POS terminal via NFC. This HCE service can be part of a mobile application with a user interface, such as a mobile wallet for payment. HCE can be used to support both closed-loop and open-loop payments. In case of open-loop NFC with HCE, it should be noted that NFC application would run on the mobile device OS rather than SE.

For open-loop NFC payment with HCE, the new EMVCo specification compliant tokens [4], typically made valid for limited period or number of use through configuration, is used by this HCE service for payment [4]. The tokens can be stored in the application itself or they could be stored in other secure locations such as a trusted execution environment (TEE) [21] or a Secure Element(SE) or alternatively, the HCE service could connect in real-time or at given intervals with a back-end server in the cloud to retrieve payment token to exchange with the contactless terminal. Since TEE and SE limited adoption of NFC until now, until introduction of HCE, NFC payment did not gain significant momentum so far. Real-time retrieval of tokens from the cloud at the moment of tapping on a reader or POS terminal is a possible but unlikely option, as network latency may result in a poor user experience. Limited validity payment token specified by EMVCo in 2014 [4], would be used in place of a physical payment card for generation and/or processing of EMV payment cryptogram. Visa, Mastercard, and Europay, together known as EMVCo, published a new specification for Payment Tokenisation in Q1, 2014. Tokenisation is a way of not exposing the permanent account number (PAN) of the cardholder. NFC payment is perhaps the principal driver of the new EMVCo tokenisation specification.

Here are some salient features of the new EMVCo payment tokenisation specification:

- The specification requires the token format to be similar to credit card numbers (13-19 digits) and comply with LUHN algorithm.
- Unlike financial tokens used today and which follow PCI guidelines for non-payment token used for analytics/reporting, the new specification compliant token can be used to initiate payments.
- The new specification compliant tokens are merchant or payment network specific, so they are only relevant to a specific domain.
- For most use cases, the PAN remains private between issuer and customer. The token becomes a payment object shared between merchants, payment processors, the customer, and possibly others within the domain.
- There is an identity verification, i.e., authentication, process to validate the requestor of a token each time a token is requested.
- The type of token generated is variable and is based on risk score.
- When tokens are used as a payment objects, there are "Data Elements" which include payment network data, cryptographic nonce and token assurance level.
- There are facilities and features to address PAN privacy, mobile payments, re-payments, EMV/smartcard, and even card-not-present web transactions.

### D. Review of Apple Pay and Tokenisation [5]

In October, 2014, with iPhone 6, Apple introduced secure contactless or NFC payment [5]. With Apple Pay, during registration, user adds actual credit and debit card numbers to backend called Passbook and a unique Device Account Number is assigned, encrypted, and securely stored in the Secure Element, a dedicated chip in iPhone. These numbers are not stored on Apple servers. And when the user makes a purchase, the Device Account Number, along with a transaction-specific dynamic, one-time-use security code called token, is used to process payment. Actual credit or debit card numbers are never shared by Apple with merchants or acquirers, or transmitted over the network. As for tokenisation in Apple Pay, tokenisation and de-tokenisation operations take place at card networks like Visa and American Express and not at processor or payment gateway. Although compliance of this tokenisation with new EMVCo tokenisation [4] is not in public domain, the authors believe that at least conceptually, these approaches are similar.

## III. CURRENT WORK - ANALYSIS OF APPLICABILITY OF ISO 9564 PIN AUTHENTICATION TO CLOSED-LOOP MOBILE PAYMENT SYSTEM

### A. ASSUMPTIONS

It is assumed that we do not want to couple the solution of securing the mobile payment authentication with the SIM (UICC) and other form factors of Secure Element [24] provided by the telecom operator or other parties.Furthermore, m-PIN for authentication may be used for various access bearers or channels like USSD, SMS, NFC (near field communication) and mobile handset apps (using HTTP(S) channel). Additional mechanisms may also be applied to secure the authentication even further.

### B. ANALYSIS

Even for open-loop NFC payment, authentication of cardholder with the help of token by the card issuer need not comply with ISO 9564 PIN standard at all and there is no strong case of supremacy of this ISO specification in case of mobile payment in general. Therefore, we focus our attention to closed-loop payment. Let us analyse two distinct scenarios.

### I. Can ISO 9564 PIN Management be extended to Closed-Loop Mobile Payment Authentication in Mobile-Device-Present Transaction?

m-PIN is entered on the user-owned mobile device for mobile-device-present transaction for in-store purchase. The merchant has an SVA in the same closed-loop system as the subscriber. The application on mobile device cannot establish truly secure cryptographic relationship with the acquirer (i.e., the backend services in case of closed-loop system) to enable PIN block encryption for online or real-time authentication which involves verification by the issuer (i.e., the backend services in case of closed-loop system) of m-PIN. Therefore it is not a good idea to think about complying with ISO 9654 for this type of use of m-PIN. Most end-user mobile devices in circulation in the market would not even pass FIPS 140-2 Level 1 certification and question of storing sensitive data securely in mobile devices does not arise.

### II. Can ISO 9564 PIN Management be extended to Closed-Loop Payment Authentication involving merchant POS terminal or Bank ATM Machine, i.e., Mobile-Device-Not-Present Transaction?

a. Because of challenge in inter-operability, only ISO 9564 PIN block format-1 may be considered (plain text m-PIN plus some filler) for online m-PIN verification by the SVA system acting as SVA issuer, for purchase on merchant web site. Of course, encrypted communication channel (SSL/TLS) may be used, but message or data level encryption of PIN block is not possible. If we have to consider other formats, MSISDN (Mobile Station International Subscriber Directory Number or mobile number) may be used as a substitute for PAN (permanent account number in case of card) and this along with m-PIN may be used for formation of block.

b. Closed-loop m-PIN would be exposed to the intermediaries of interfacing open-loop system (e.g., merchant POS terminal or ATM machine of the bank) because of plaintext data. Because encrypted communication channel must terminate at intermediary.

c. In card based system, PIN is used only for card-present transaction and never for card-not-present transaction. Card-not-present transaction (e.g., online shopping) uses CVV or equivalent and 3D Secure technology specified by networks or some form of OTP (one-time-password). Somebody using closed-loop SVA for in-store shopping and using merchant's POS terminal would not require using the mobile device (which provides a level of authentication –something the user has). This is equivalent to online mobile payment or mobile-device-not-present payment and use of m-PIN only for authentication for such use case, is not a good idea.

d. Therefore it is not a good idea to think of compliance with ISO 9654 for this type of use of m-PIN.

## C. FUTURE DIRECTION OF CLOSED-LOOP MOBILE PAYMENT AUTHENTICATION

In closed-loop payment system, the issuer of stored-value-account (SVA) is typically a telecom operator and in some cases, a bank. The acquirer of transaction where SVA is used as a payment instrument for a purchase on a merchant site or in store, is a POS terminal or ATM machine or a payment gateway of web site. Specific customer authentication requirements for such systems are not standardised.

On the other hand, open-loop card-based payment uses card PIN for card-present payment authentication and this is based on ISO 9564 standard. Visa and Mastercard based open-loop payment schemes or networks with associated issuers, acquirers, merchants and cardholders are very well known.

Most mobile devices in circulation in the market would not even pass FIPS 140-2 Level 1 certification and question of storing sensitive data securely in mobile devices does not arise. Although Trusted Computing Group [22] (TCG)'s Mobile Trusted Computing (MTM) [22] or GlobalPlatform's Trusted Execution Environment (TEE) [21] are approaches for making mobile devices more secure, adoption of such specification in the market is insignificant.

Besides m-PIN, out-of-band or in-band OTP, device fingerprinting, etc may be used to strengthen authentication in closed-loop mobile payment. There is no chance that people would talk about applying ISO 9564 standard to this use case.

## V. CONCLUSION

Card professionals who are coming to the domain of mobile payments often try to apply ISO 9564 card PIN authentication to mobile payment and even go to the extent of complying with this standard for closed-loop mobile payment. The authors highlighted that such compliance is misplaced. Instead of such compliance, we must use standard channel level encryption for direct transmission of m-PIN between mobile device and the remote closed-loop payment service and other mechanisms for stronger authentication. Question of applicability of ISO 9564 is often raised by card professionals in mobile payments domain and this has not been addressed by anybody so far in the payment security industry, as per knowledge of the authors. The authors make categorical statement about applicability of this ISO standard and hope that this analysis would help both card professionals coming to mobile payments and also existing mobile payment professionals to take informed decision regarding security and compliance.

Since closed-loop mobile payment is a relatively new field, the authors make a case for Common Criteria Recognition Agreement (CCRA) [3] or other standards body to push for publication of a mobile device-agnostic Protection Profile or specific standard, respectively, incorporating the suggested high-level requirements (e.g., m-PIN, OTP, device fingerprinting, etc).

## REFERENCES

**[1]** ISO (International Organisation for Standardisation), Standard 9564-1: 2011, Financial services -- Personal Identification Number (PIN) management and security -- Part 1: Basic principles and requirements for PINs in card-based systems - http://www.iso.org/iso/home/store/catalogue_tc/catalogue_detail.htm?csnumber=54083
**[2]** ISO 9564 Standard, Wikipedia, http://en.wikipedia.org/wiki/ISO_9564
**[3]** Common Criteria - https://www.commoncriteriaportal.org - The CC is the driving force for the widest available mutual recognition of secure IT products worldwide.

http://www.emvco.com/specifications.aspx?id=263

**[5]** Apple Pay Contactless Secure Payment and Tokenisation - https://www.apple.com/iphone-6/apple-pay/

**[6]** Fraud Protection for Native Mobile Applications, ThreatMetrix TrustDefender Mobile, http://www.threatmetrix.com/wp-content/uploads/2014/11/TrustDefender-Mobile-Technical-Brief.pdf

**[7]** Host Card Emulation (HCE) Whitepaper by Smartcard Alliance - http://www.smartcardalliance.org/wp-content/uploads/HCE-101-WP-FINAL-081114-clean.pdf

**[8]** Future of Secure Mobile Payments by Amal Saha, CISO Platform Annual Summit, 2013 - http://www.slideshare.net/cisoplatform7/future-of-secure-mobile-payments-amal-saha , http://www.youtube.com/watch?v=6xfIkLKWlko

**[9]** Google Host Card Emulation:https://developer.android.com/guide/topics/connectivity/nfc/hce.html

**[10]** Payment Card Industry PIN Security Requirements, Version 1.0, September 2011, PCI Security Standards Council

**[11]** FIPS PUB 140–2: Security Requirements for Cryptographic Modules, Federal Information Processing Standard, latest version published on May 25, 2001

**[12]** EMVCo Contact and Contactless Terminal Approval - Level 1 and Level 2 - http://www.emvco.com/approvals.aspx?id=56

**[13]** Device Fingerprinting in mobile payment use case - IBM Trusteerhttp://www.trusteer.com/products/trusteer-pinpoint-criminal-detection

**[14]** Sugata Sanyal, Ayu Tiwari and Sudip Sanyal. "A multifactor secure authentication system for wireless payment." Emergent Web Intelligence: Advanced Information Retrieval. Springer London, 2010.341-369.

**[15]** Ayu Tiwari, Sudip Sanyal, Ajith Abraham, S. J. Knapskog and Sugata Sanyal, (2011). A multifactor security protocol for wireless payment-secure web authentication using mobile devices.arXiv preprint arXiv:1111.3010.

**[16]** Hristo Bojinov et al. "Mobile Device Identification via Sensor Fingerprinting." arXiv preprint arXiv:1408.1416 (2014).

**[17]** Michael Rausch, Nathan Good, and Chris Jay Hoofnagle. "Searching for Indicators of Device Fingerprinting in the JavaScript Code of Popular Websites.", Proceedings, Midewest Instruction and Computing Symposium, 2014.

**[18]** M Rausch, A Bakke, S Patt, B Wegner and D Scott. Demonstrating a Simple Device Fingerprinting System, Proceedings, Midewest Instruction and Computing Symposium, 2014.

**[19]** Akli Adjaoute, Multi-Dimensional Behavior Device ID, US Patent Publication Number US20140164178 A1, http://www.google.com/patents/US20140164178

**[20]** Services in Android Application Framework - http://developer.android.com/guide/components/services.html

**[21]** GlobalPlatform's Trusted Execution Environment (TEE), http://www.globalplatform.org/specificationsdevice.asp

**[22]** Trusted Computing Group (TCG), http://www.trustedcomputinggroup.org and http://www.trustedcomputinggroup.org/solutions/mobile_security

**[23]** M-PESA, http://en.wikipedia.org/wiki/M-Pesa - M-Pesa (M for mobile, pesa is Swahili for money) is a mobile-phone based money transfer and microfinancing service, launched in 2007 by Vodafone for Safaricom and Vodacom, the largest mobile network operators in Kenya and Tanzania

**[24]** Secure Element and smart card form factors as per GlobalPlatform, http://globalplatform.org/mediaguideSE.asp